# Evaluating the Efficiency and Cost-effectiveness of RPB-based $CO_2$ Capture: A Comprehensive Approach to Simultaneous Design and Operating Condition Optimization

Howoun Jung[a], Nohjin Park[b], Jay H. Lee[a,*]

[a] Mork Family Department of Chemical Engineering and Materials Science, University of Southern California, Los Angeles, CA 90089, USA (Corresponding author: jlee4140@usc.edu)

[b] GS Engineering and Construction Corp., Carbon Solution Research Team/RIF Tech, Seoul, Republic of Korea

**Abstract**

Despite ongoing global initiatives to reduce $CO_2$ emissions, implementing large-scale $CO_2$ capture using amine solvents is fraught with economic uncertainties and technical hurdles. The Rotating Packed Bed (RPB) presents a promising alternative to traditional packed towers, offering compact design and adaptability. Nonetheless, scaling RPB processes to an industrial level is challenging due to the nascent nature of its application. The complexity of designing RPB units, setting operating conditions, and evaluating process performance adds layers of difficulty to the adoption of RPB-based systems in industries. This study introduces an optimization-driven design and evaluation for $CO_2$ capture processes utilizing RPB columns. By employing detailed process simulation, we aim to concurrently optimize unit design and operating parameters, underscoring its advantage over conventional sequential approaches. Our process design method integrates heuristic design recommendations as constraints, resulting in 9.4% to 12.7% cost savings compared to conventional sequential design methods. Furthermore, our comprehensive process-level analysis reveals that using concentrated MEA solvent can yield total cost savings of 13.4% to 25.0% compared to the standard 30wt% MEA solvent. Additionally, the RPB unit can deliver an 8.5 to 23.6 times reduction in packing volume. While the commercial-scale feasibility of RPB technology has been established, the advancement of this field hinges on acquiring a broader and more robust dataset from commercial-scale implementations. Employing strategic methods like modularization could significantly reduce the entry barriers for $CO_2$ capture projects, facilitating their broader adoption and implementation.

---

## 1. Introduction

Capturing $CO_2$ from fossil-based $CO_2$ emission sources is pivotal for mitigating climate change impacts and transitioning smoothly to renewable resources by 2050 [1, 2]. Post-combustion $CO_2$ capture with amine solvents is a well-established technology, as demonstrated by multiple industrial-scale facilities worldwide [3]. Yet, lessons from major projects like Petra Nova reveal large-scale operations' economic and technical challenges [4]. Despite the potential benefits of economies of scale, the substantial upfront investment and risks make massive $CO_2$ capture plants less appearing financially. While $CO_2$ capture is an essential component of the energy transition, it should not be viewed as the ultimate goal. To promote broader adoption of $CO_2$ capture, focusing on smaller and medium-scale processes might be more practical, offering templates for diverse applications [5]. Process intensification, which aims to boost efficiency and applicability through advanced techniques, fits well within this approach.

The Rotating Packed Bed (RPB), a type of HiGee (high gravity) unit, offers a promising solution. Pioneered by Ramshaw and Mallinon [6], RPB enhances mass transfer through rapid rotation and centrifugal acceleration, creating a gravitational field 100-1000 times stronger. This enhancement dramatically improves mass transfer and processing throughput [7, 8], thanks to the formation of thinner liquid films and smaller droplets that expand the effective surface area for mass transfer [9]. Additionally, the high centrifugal force allows for the use of packing materials with greater specific surface areas, thereby addressing the typical challenges of flooding and pressure drops encountered in traditional columns [10].

Recognizing RPB's potential to drastically reduce the size and cost of mass transfer units, researchers are exploring its application in $CO_2$ capture, which typically accounts for a substantial portion of capital expenditure (CAPEX), typically comprising 30-50% [11]. Diverse studies have investigated the integration of RPB with various solvents such as monoethanolamine (MEA), piperazine (PZ), and diethylenetriamine (DETA). Notably, Jassim [12] focused on pilot-scale RPB absorbers with MEA solvent, highlighting the importance of amine concentration and inlet $CO_2$ levels. Meanwhile, Yu et al. [13] compared four different amine solvents (MEA, DETA, MEA+PZ, DETA+PZ), identifying DETA as particularly promising for RPB-based capture due to its high absorption capacity and fast kinetics. Wu et al. [14] demonstrated that using a PZ+DETA blend with RPB could cut regeneration energy by about 45% compared to the standard 30wt% MEA. By offering a detailed exploration of RPB's efficiency and potential for $CO_2$ capture, these studies contribute to a more nuanced understanding of how to make $CO_2$ capture technology more feasible and financially viable on different scales.

Quantitative assessments through simulation studies of RPB-based $CO_2$ capture processes highlight the benefits of integrating RPB units [15-17]. Chamchan et al. [5] demonstrated that using 30wt% MEA in RPB units can achieve comparable $CO_2$ capture and energy consumption while requiring only 35% of the volume. Similarly, Im et al. [18] explored a pilot-scale process with MEA solvent, finding that RPB absorbers could reduce packing volume by about 30%. Despite numerous experimental and simulation studies demonstrating RPB's versatility, its extrapolation to a commercial scale is tentative and remains somewhat scarce. There is a noticeable lack of systematic process-level comparisons with RPB-based $CO_2$ capture processes and traditional fixed packed beds (PB) methods.

Recent research has delved into modeling and cost analysis of a commercial-scale RPB-based $CO_2$ absorber unit, specifically targeting around 2200 tons per day (TPD) scale using 55 and 75wt% MEA solvents [19]. Despite the apparent benefits of scaling up, the suitability of RPB units for large-scale applications is questionable due to their inherent design for small scales dictated by rotational dynamics. In addition, the lack of large-scale RPB implementation experience hampers the ability to apply findings to larger, practical settings. Although Agarwal et al. [10] introduced a systematic RPB column design procedure, it is primarily based on lab-scale heuristics and mainly addresses basic design necessities to avoid flooding. Scaling this procedure directly to larger RPB designs results in impractical or inefficient designs due to substantial pressure drops. Moreover, RPB design using this procedure depends heavily on initial assumptions about operational variables, such as rotation speed, complicating the discovery of a global optimum for large-scale configurations and operating conditions. A lack of direct comparison with processes using 30wt% MEA further complicates the evaluation of its suitability for large-scale use. Furthermore, there is a noticeable gap in research regarding the optimal design and operation of RPBs within integrated carbon capture systems, highlighting a significant opportunity for future research and industrial application.

To tackle the noted challenges, this study adopts a simultaneous optimization approach for designing RPB units and their operating conditions. By coupling cost-based optimization with detailed modeling and techno-economic analysis (TEA), we aim not only to reduce expenses but also to pinpoint the specific decisions that lead to cost savings. In addition, we integrate heuristics-based design recommendations as constraints within the optimization framework. This integration yields more reliable and practical designs for larger-scale RPB units, which are not effectively addressed by purely heuristic-based approaches. The key contributions of this work can be summarized as follows:

1. Extensive Feasibility Investigation: This research extensively investigates the viability of commercial-scale RPB-based $CO_2$ capture processes. It encompasses process modeling, design and operating parameters optimization, and techno-economic analysis (TEA). This comprehensive approach addresses a significant gap in existing research, as no previous studies have integrated both optimization and TEA for RPB-based $CO_2$ capture systems.

2. Innovative Design Approach for Commercial-Scale RPB Processes: The research introduces a novel design strategy for RPB processes applicable to *large-scale* operations. Traditional heuristic-based designs often fall short when scaled up, leading to impractical or inefficient outcomes. Our approach, grounded in first principles and optimization, offers a thorough and effective scale-up of RPB processes. This method is particularly adaptable; not only does it accommodate existing heuristic insights, but it can also seamlessly integrate new findings or design requirements. The flexibility of optimization problems in efficiently managing constraints makes this approach especially valuable for evolving design needs.

3. Advancement in RPB-based Process Optimization: This study introduces a cost-effective $CO_2$ capture process using RPB, achieved by optimizing both the design and operational conditions, and includes a TEA for commercial-scale applications. To the best of our knowledge, this represents the first initiative in optimizing a commercial-scale RPB-based $CO_2$ capture process. Including design parameters as decision variables in the optimization process not only enhances the overall strategy but also significantly broadens the scope of knowledge in the field of RPB processes. This approach marks a pivotal step in advancing the practicality and efficiency of RPB technology for large-scale $CO_2$ capture.

4. Systematic Comparative Analysis: Our research includes a detailed comparison between RPB units and conventional fixed PB processes. This comparative analysis aims to clearly outline both the advantages and limitations of RPB units. The study goes beyond merely identifying the benefits of using RPB units and specific solvents; it delves into the underlying reasons for these advantages. By doing so, it provides a nuanced understanding of the value proposition of RPB technology, offering clear insights into how and why RPB units can be a superior choice in specific contexts. This aspect of the study is crucial for stakeholders to make informed decisions regarding the adoption of RPB-based $CO_2$ capture processes.

The subsequent sections of this paper are structured as follows: Section 2 presents the mathematical model for the RPB-based $CO_2$ capture process. Section 3 elucidates the unit design and scale-up approaches employed in this study. The scale-up results and accompanying discussions are detailed in Section 4. Finally, Section 5 encapsulates the conclusions derived from this research.

## 2. System Description

Figure 1 illustrates the schematic diagram of a typical RPB unit. The RPB comprises a rotor, an annular packed bed positioned between two disks on a rotating shaft, all enclosed in a casing.

The mass transfer process is driven by the co-current or counter-current radial flow of liquid and gas through the packing rotor, making the rotor's design a critical factor influencing pressure drop, flooding conditions, and overall mass transfer efficiency.

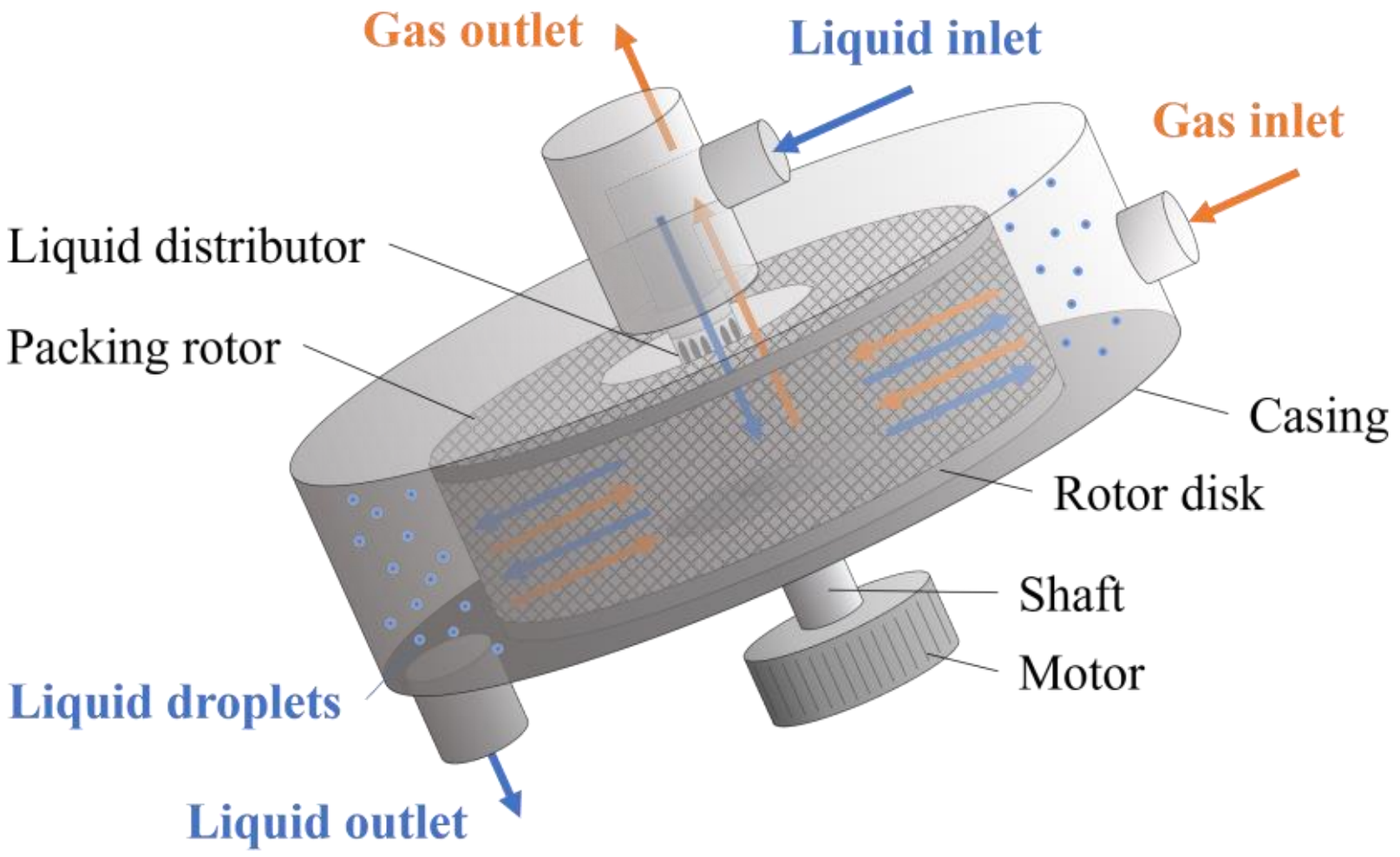

Figure 1 Schematic diagram of counter-current Rotating Packed Bed (RPB)

The RPB's versatility is reflected in its various rotor configurations, including basic packed beds, dual motors, zigzag baffles, packing-filled designs, and spraying beds, each catering to different operational needs and scenarios [20]. In the specific counter-current basic bed configuration considered in this study, gas is drawn radially inward from the perimeter, propelled by a pressure gradient, while liquid is forced radially outward from the center due to centrifugal acceleration. This dynamic interaction results in the formation of thinner liquid films and smaller droplets upon contact with the packing material, thereby enhancing the mass transfer rate. Additionally, the centrifugal acceleration imposed by the RPB allows for a significantly higher flooding threshold than conventional PBs, permitting the use of packing materials with a higher specific surface area, like wire mesh packings, to further optimize mass transfer.

## 2.1. Process model

### 2.1.1. Thermodynamic and physical properties

In the RPB-based $CO_2$ capture process, concentrated amine solvents are preferred due to their effectiveness in high-viscosity environments. MEA, a reference solvent, is frequently used, with concentrations typically up to 75% [18, 19, 21]. To capture the full spectrum of operational conditions, our study examines a wide range of MEA concentrations, from 30 to 70wt%. We employ the eNRTL thermodynamic model to understand this extensive range's chemical and phase equilibria. This model incorporates updated parameters calibrated explicitly for the MEA concentrations under consideration [22], providing a comprehensive and accurate representation of chemical and phase behaviors at various concentrations, as depicted in Figure 2.

(a) 30wt% MEA (b) 60wt% MEA

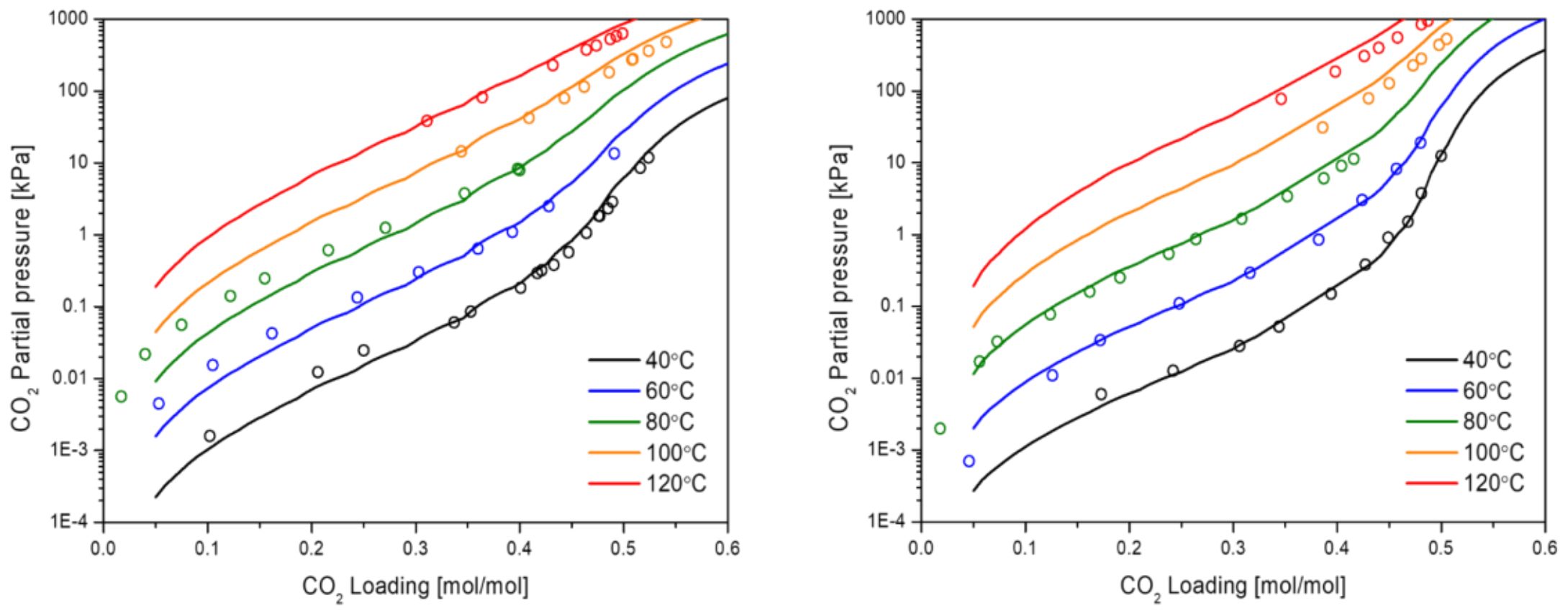

Figure 2 $CO_2$ partial pressure of MEA solvent (a) 30wt% and (b) 60wt% (solid line: model; circle: experiment [23])

We employed the physical property models detailed in Table 1. While no model can assure absolute accuracy across an extensive range of concentrations, we prioritized employing models that have been validated through experimental methods over those based on estimations.

Table 1 Thermodynamic and physical property models for MEA-$H_2O$-$CO_2$ system

| Properties | Model |
|---|---|
| Vapor-liquid equilibrium | eNRTL [22] & ideal gas |
| Liquid density | Jayarathna et al. [24] |
| Liquid specific heat capacity | Agbonghae et al. [25] |
| Liquid viscosity | Amundsen et al. [26] |
| Liquid diffusivity | Ko et al. [27] / Versteeg et al. [28] |
| Gas specific heat capacity | DIPPR from Aspen plus |
| Gas diffusivity | Fuller et al. [29] |
| Heat of absorption | Clausius-Clapeyron relation |

### 2.1.2. RPB column model

Modeling the RPB requires a thorough understanding of the radial flow pattern and how rotation influences the mass transfer coefficient and surface area. To capture the critical role of mass transfer phenomena, our RPB column model employs a rate-based approach grounded in the two-film theory for calculating mass and heat transfers. This is combined with an enhancement factor model to account for reaction-enhanced transfers. Given that both vapor and liquid phases flow counter-currently along the RPB units' radial axis, the model is designed with a radial distribution domain ($r$), as depicted in Figure 3.

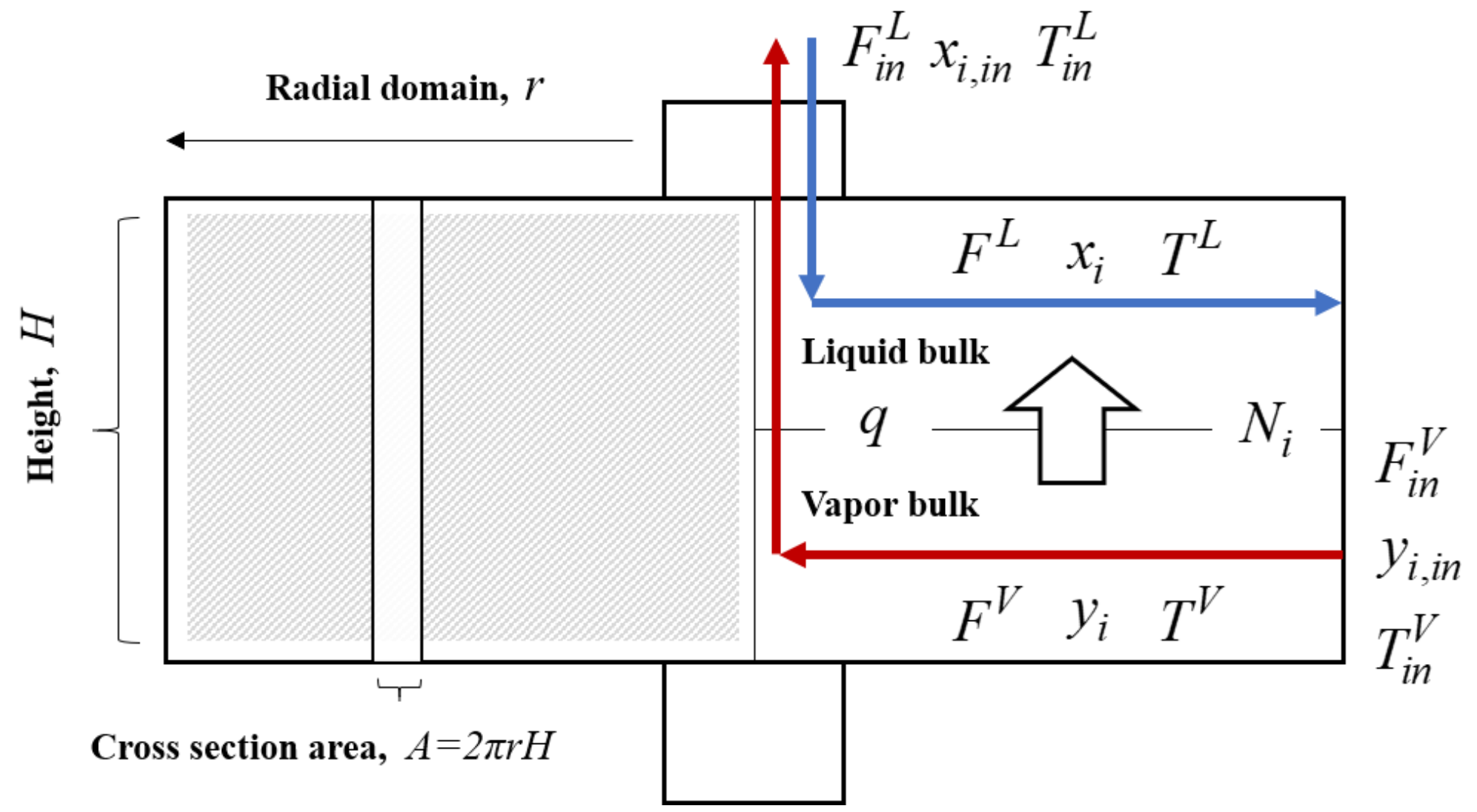

Figure 3 Conceptual scheme of RPB column model

For the RPB column model, the following assumptions are made:

1) The flow is one-dimensional in the radial direction.
2) Vapor and liquid phases are distinct, and each phase is considered well-mixed.
3) Vapor and liquid flows are treated as plug flows, neglecting any dispersion effects.
4) An enhancement factor is included to reflect the enhanced mass transfer due to chemical reactions.
5) The model assumes no heat loss or degradation of the amine solvent.

Under these assumptions, the mass balance equations for the RPB are formulated as partial differential equations (PDEs) with the following structure [18]:

For vapor phase,

$$\varepsilon^V \frac{\partial C_i^V}{\partial t} = \frac{1}{2\pi r H}\frac{\partial (F^V y_i)}{\partial r} - a^I N_i \tag{1}$$

$$B.C.\quad \frac{\partial (F^V y_i)}{\partial r} = 0 \quad at\, r = r_{inner}\ \&\ \ F^V y_i = F_{in}^V y_{i,in}\ at\, r = r_{outer}$$

For liquid phase,

$$\varepsilon^L \frac{\partial C_i^L}{\partial t} = -\frac{1}{2\pi r H}\frac{\partial (F^L x_i)}{\partial r} + a^I N_i$$

$$B.C.\quad F^L x_i = F_{in}^L x_{i,in}\ at\, r = r_{inner}\ \&\ \ \frac{\partial (F^L x_i)}{\partial r} = 0\ at\, r = r_{outer} \tag{2}$$

Here, $r_{inner}$ and $r_{outer}$ represent the inner and outer radii of an RPB unit. The continuity equations of $F = 2\pi r z C_{tot} u$ for vapor and liquid phases are expressed considering the changing control volume along the radial axis. Similarly, the energy balances are established:

For vapor phase:

$$\varepsilon^V C_{tot}^V C_p^V \frac{\partial T^V}{\partial t} = \frac{F^V C_p^V}{2\pi r H}\frac{\partial T^V}{\partial r} - a^I h^I (T^L - T^V) \tag{3}$$

$$B.C.\ \frac{\partial T^V}{\partial r} = 0 \ \ at\, r = r_{inner}\ \&\ \ T^V = T_{in}^V \ \ at\, r = r_{outer}$$

For liquid phase:

$$\varepsilon^L C_{tot}^L C_p^L \frac{\partial T^L}{\partial t} = -\frac{F^L C_p^L}{2\pi r H}\frac{\partial T^L}{\partial r} + a^I\left(h^I(T^V - T^L) + N_{H_2O}\Delta H_{H_2O}^{vap} + N_{CO_2}\Delta H_{CO_2}^{abs}\right)$$

$$B.C.\ \ T^L = T_{in}^L \ \ \text{at}\, r = r_{inner}\ \&\ \frac{\partial T^L}{\partial r} = 0 \ \ \text{at}\, r = r_{outer} \tag{4}$$

These equations incorporate the heat of absorption from $CO_2$ and the heat of evaporation of water on the liquid side, while conductive heat transfer is considered between both phases. The absorption heat is estimated from the $CO_2$ equilibrium diagram with the Clausius-Clapeyron relation.

#### 2.1.3. Mass and heat transfer model

Utilizing the two-film theory, we calculate the mass transfer rate using the overall mass transfer coefficient $K_i^{overall}$ and the phase equilibrium driving force. The overall mass transfer coefficient reflects the mass transfer resistance in both vapor and liquid phases. Our RPB column model incorporates the enhancement factor model for the liquid side $CO_2$ mass transfer while only considering vapor resistance for other components:

$$N_i = K_i^{overall}(P_i - P_i^*) \tag{5}$$

For $CO_2$:

$$K_i^{overall} = \frac{1}{\frac{RT^V}{k_i^V} + \frac{He_i}{E_i \cdot k_i^L}} \tag{6}$$

For other components:

$$K_i^{overall} = \frac{k_i^V}{RT^V} \tag{7}$$

The enhancement factor model for the irreversible 2nd order reaction is determined using [30]:

$$E_{CO_2} = 1 + ((E_1 - 1)^{-1.35} + (E_2 - 1)^{-1.35})^{-\frac{1}{1.35}}$$

$$\text{where } E_1 = \frac{Ha}{tanh(Ha)} \quad \text{and} \quad E_2 = 1 + \frac{D_{MEA} C_{MEA}}{2 D_{CO_2} C_{CO_2}}, \qquad Ha = \frac{\sqrt{k_{app} D_{CO_2}}}{k^L} \tag{8}$$

For the reaction kinetics ($k_{app}$), we adopt concentration-based reaction kinetics with a direct reaction mechanism from Luo et al. [31], suitable for various MEA concentrations.

In the RPB model, incorporating the effect of the packing's rotation is vital for capturing the true dynamics of RPB systems and ensuring the model's validity. We calculate the vapor and liquid mass transfer coefficients for the RPB using established correlations from Onda et al. [32] and Tung et al. [33], respectively, with an adaptation to replace the gravitational force ($g$) with the centrifugal force ($\omega r^2$), where $\omega$ is the angular velocity and $r$ the radius from the rotation axis. For the vapor phase mass transfer coefficient ($k^V$), the formula is adapted as follows:

$$k^V = 2.0 \cdot \left(\frac{\rho^V d_h u^V}{\mu^V}\right)^{0.7} \left(\frac{\mu^V}{\rho^V D^V}\right)^{1/3} \left(\frac{D^V}{a_p d_p^2}\right) \tag{9}$$

$$\frac{k^L d_p}{D^L} = 0.918 \cdot \left(\frac{\rho^L d_h u^L}{\mu^L}\right)^{1/3} \left(\frac{\mu^L}{\rho^L D^L}\right)^{1/2} \left(\frac{d_h^3 {\rho^L}^2 \omega r^2}{{\mu^L}^2}\right)^{1/6} \left(\frac{a_p}{a}\right)^{1/3} \tag{10}$$

The effective specific surface area ($a$) also includes centrifugal force terms as follows:

$$\frac{a}{a_p} = C_{cavity} \left(\frac{202.3}{546.5}\right) \left(\frac{\omega r^2}{g_0}\right)^{0.0435} \left(\frac{u^L}{u_0^L}\right)^{0.4275} \left(\frac{\mu}{\mu_0}\right)^{0.12} \left(\frac{\gamma}{\gamma_0}\right)^{-0.5856} \tag{11}$$

The effective surface area is calculated from the modified correlation from Xie et al. [34], considering operational ranges of 30 to 90wt% MEA. It includes a correction factor of 1.15, accounting for an additional 15% surface area contribution from the cavity zone of the RPB unit [35].

### 2.1.4. Hydraulic model

In our RPB model, we have incorporated models for pressure drops and liquid holdup to describe the hydraulic behavior within the RPB units. The differential pressure drop model, based on the work of Llerena-Chavez [36], is applied as follows:

$$\frac{\partial P}{\partial r} = \frac{150(1-\varepsilon)^2\mu^V}{d_p^2\varepsilon^3}\left(\frac{G}{2\pi Hr}\right) + \frac{1.75(1-\varepsilon)\rho^V}{d_p\varepsilon^3}\left(\frac{G}{2\pi Hr}\right)^2 + \rho_g\omega^2 r \tag{12}$$

This model encompasses pressure drop considering contributions from Ergun-like drag force (the first two terms) and centrifugal force (the last term). We have deliberately omitted gas-solid slip correction factors to avoid over-extending the model beyond its validated lab-scale conditions. For estimating liquid holdup, we employ the correlation from Burn's research [9]:

$$\varepsilon^L = 0.039\left(\frac{\omega r^2}{g_0}\right)^{-0.5}\left(\frac{u^L}{u_0^L}\right)^{0.6}\left(\frac{v}{v_0}\right)^{0.22} \tag{13}$$

Here, $g_0$, $u_0^L$ and $v_0$ represent characteristic centrifugal acceleration, flow rate, and kinematic viscosity, respectively.

### 2.1.5. Overall process model

Figure 4 displays the comprehensive flowsheet of the RPB-based $CO_2$ capture process, encompassing the developed RPB column model alongside other necessary units within the gPROMS simulation environment. In this standard amine-based $CO_2$ capture configuration, flue gas enters the absorber (ABS), where $CO_2$ is absorbed, and the purified gas exits from the top of the RPB absorber. The $CO_2$-enriched solvent from the absorber is then directed to a main heat exchanger (HX), where it recovers heat energy from the lean solvent leaving the stripper (STR). Post heat exchange, the rich solvent enters the stripper, where the exothermic nature of $CO_2$ absorption necessitates heating to regenerate the solvent. A reboiler provides this heating. The process culminates with the collection of stripped $CO_2$ from the stripper's top, while the $CO_2$-depleted solvent is circulated back to the absorber to continue the capture process.

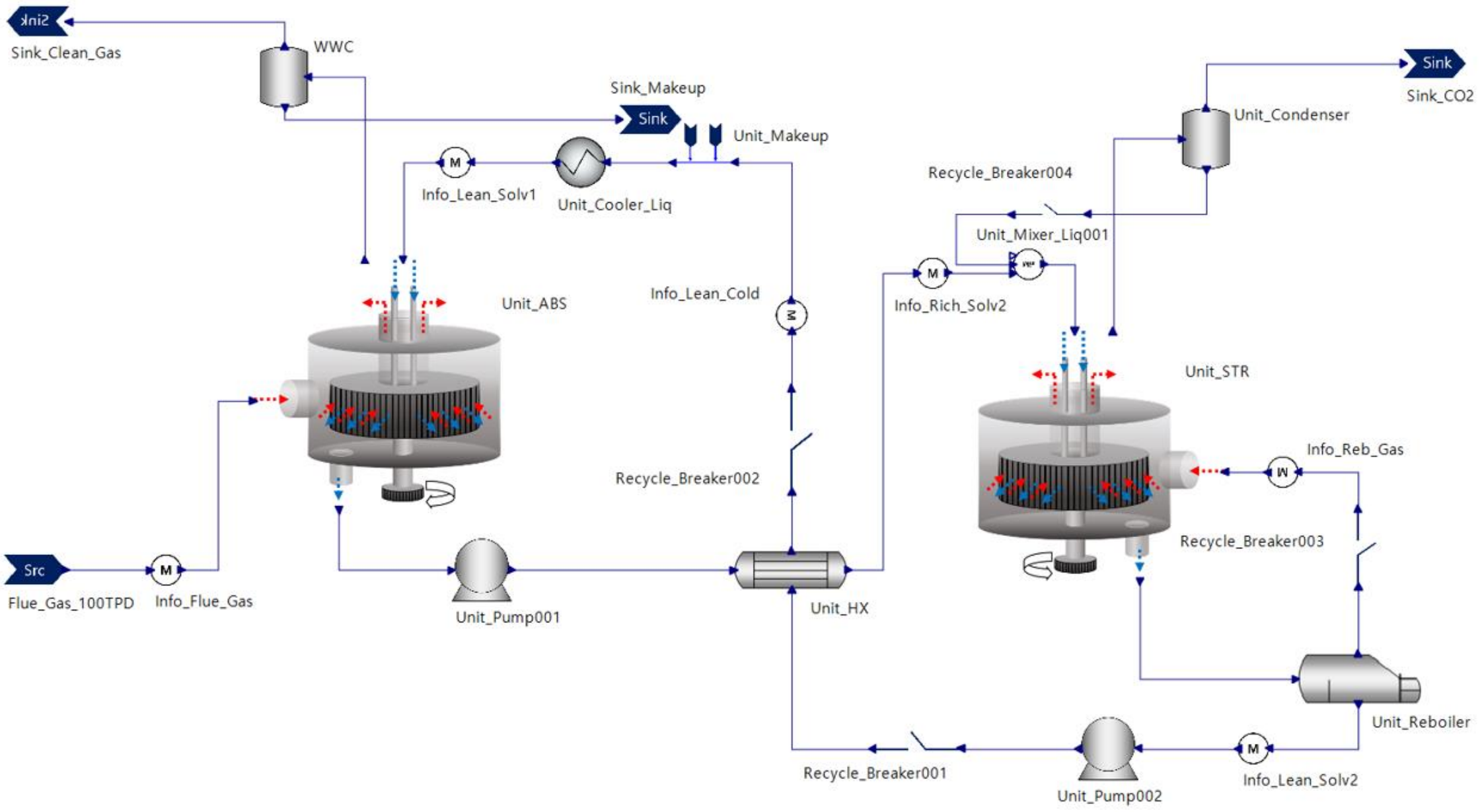

Figure 4 Overall flowsheet of RPB-based $CO_2$ capture process model (captured from gPROMS)

## 2.2. Model validation

Complete operational data on RPB-based $CO_2$ capture processes is scarce in the literature, prompting us to validate our developed RPB column model using data from pilot plant operations of separate absorbers and strippers. While several prior studies [18, 19, 21] have validated their absorber RPB models using experimental data from Jassim [12, 37], the operating conditions in Jassim's studies often deviate markedly from typical scenarios, characterized by unusually high L/G ratios (approximately 15 to 29 kg/kg) or low $CO_2$ loadings (0.05 to 0.10 mol/mol). Such discrepancies could potentially compromise model accuracy under standard operating conditions.

To enhance the relevance and reliability of our model, we have validated it using data not only from Jassim (Runs 1 to 16) but also from Kolawole [38] (Runs 1 to 36), the latter of which covers a more practical range of operating parameters (1.3 to 3.5 kg/kg L/G ratio and 0.10 to 0.22 mol/mol $CO_2$ loading), closely mirroring real-world conditions for MEA-based $CO_2$ capture. Our RPB model demonstrates mean absolute relative errors (MARE) of 11.4% for $CO_2$ capture rate and 8.8% for rich solvent temperature, maintaining ±10%p and ±20% errors for all considered conditions. Additionally, pilot plant data from Cheng et al. [39] (Runs 1 to 5 & 8 to 12) was used to validate the stripper column of the RPB model. Runs 6 and 7 were excluded from the validation due to their pinch conditions and large errors. The MARE for reboiler duty was 9.9%, with a 5.2% error in lean loading. Figures 4 and 5 display the comparative results between our model's outputs and the experimental data for key performance indicators like $CO_2$ capture rate and reboiler duty, illustrating the model's alignment with empirical observations.

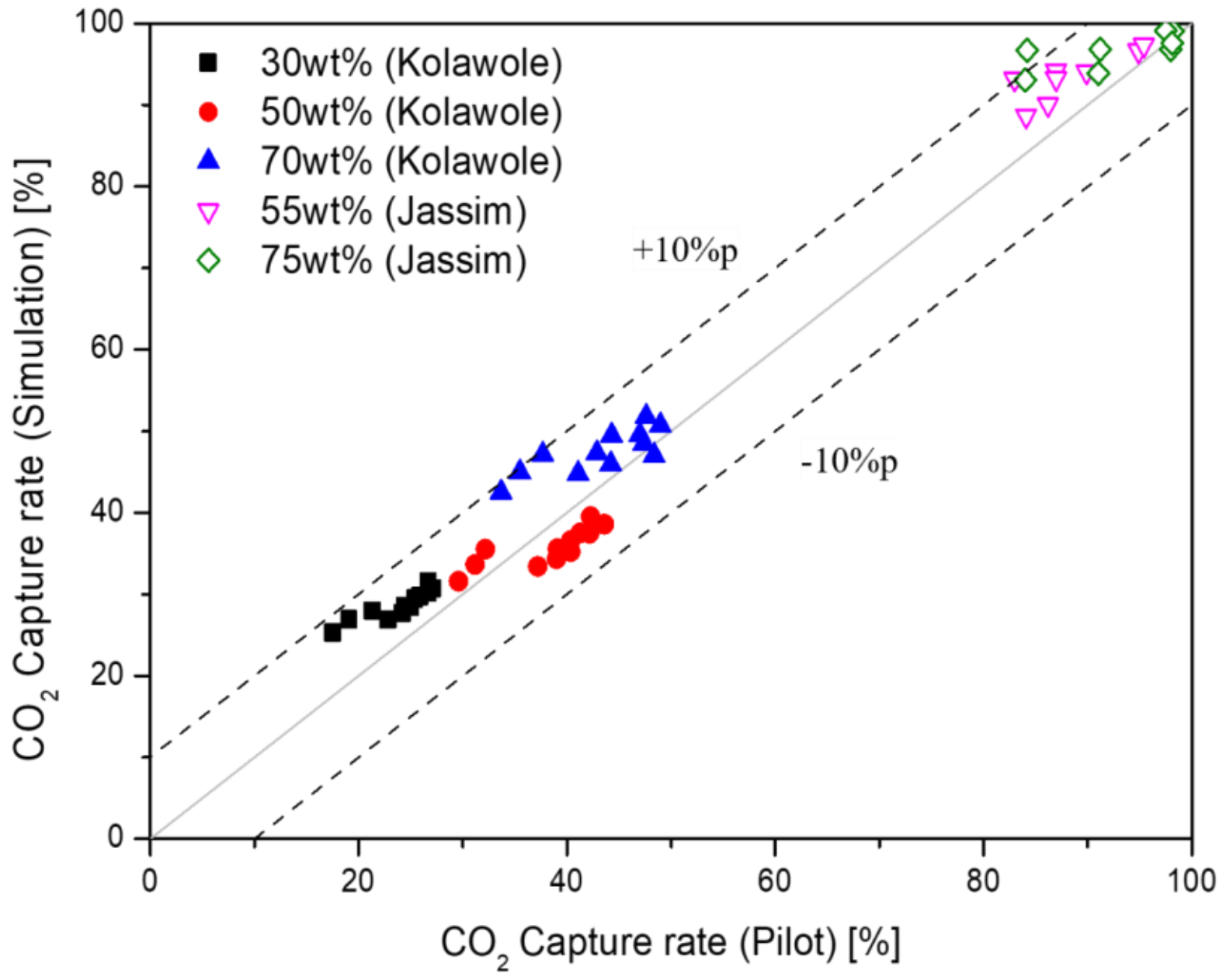

Figure 5 Comparison of model simulation and pilot plant data for RPB absorber ($CO_2$ Capture rate)

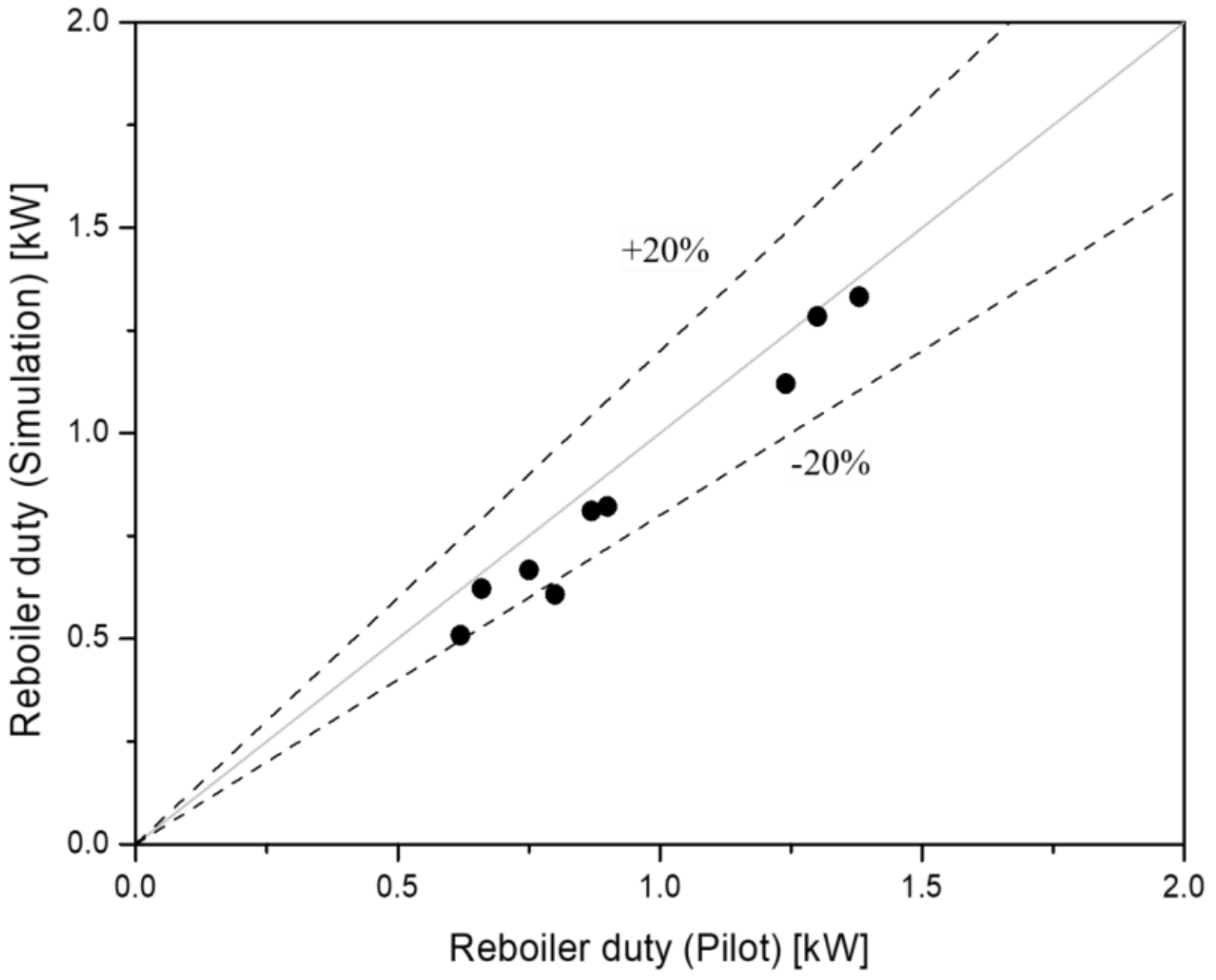

Figure 6 Comparison of model simulation and pilot plant data for RPB stripper (Reboiler duty)

# 3. Scale-up of RPB-based $CO_2$ capture process

## 3.1. Target scale and scale-up methods

An analysis of large-scale RPB-based $CO_2$ capture processes reveals inefficiencies. The energy required for rotation and the resulting pressure drop scales quadratically with the size of the unit. This is exacerbated by the need to supply momentum and centrifugal energy to rotate the liquid and vapor within the RPB, which increases with the size of the process unit. Furthermore, implementing rotational mechanics on a large scale is generally discouraged due to associated safety concerns and the complexity of maintenance challenges. Notwithstanding potential economies of scale, the practical viability of large-scale RPB units is questionable. Their characteristics seem better suited for small-to-medium-scale applications [40]. Based on this consideration, this study focuses on a small-to-medium-scale, 100 TPD (ton per day) $CO_2$ capture process for flue gas from a coal-fired power plant. The inlet composition is obtained using Mac Dowell's method [41], assuming saturated $H_2O$, and the flow rate is adjusted for a 90% $CO_2$ capture rate, matching the 100 TPD scale. This scale $CO_2$ corresponds to the flue gas flow from an approximately 6 MW power plant. The specific input stream specifications for the RPB-based $CO_2$ capture process are detailed in Table 2.

Table 2 Flue gas stream condition

| Variable | Value |
|---|---|
| Scale (TPD $CO_2$ capture) | 100 |
| Flue gas flow rate (kg/s) | 5.94 |
| Temperature (°C) | 40 |
| Pressure (bar) | 1.1 |
| Composition (mol%) | |
| - $CO_2$ | 14.5 |
| - $H_2O$ | 6.8 |
| - $N_2$ | 76.6 |
| - $O_2$ | 2.1 |

Upon establishing the input stream data for $CO_2$ capture, this study delves into two distinct process design procedures for the specified capture scale: the sequential and simultaneous design approach. The sequential design approach involves initially setting assumptions about operating conditions to determine RPB unit designs, followed by optimizing these conditions for the overall process. This step-by-step refinement of the RPB units aligns with process needs but can sometimes yield less-than-optimal results due to the interconnected nature of unit design, operating conditions, and process efficacy. In contrast, the simultaneous design approach, which we advocate, concurrently optimizes RPB unit designs and operating conditions by focusing on minimizing the total $CO_2$ capture cost. This holistic optimization potentially unveils new insights and more effective solutions that the sequential approach might overlook.

We conduct a thorough cost analysis of RPB-based $CO_2$ capture processes using both design approaches, illustrating the significant cost-saving potential of the simultaneous method.

Additionally, the study evaluates the use of 30, 50, and 70wt% MEA solvents to assess the benefits and drawbacks of employing RPB column units and concentrated MEA solvents in terms of process efficiency and total $CO_2$ capture cost. This analysis is compared against a benchmark process using 30wt% MEA solvent and PB columns, offering a comprehensive view of the implications of broader solvent choices and process design strategies.

### 3.2. Sequential design approach

#### 3.2.1. RPB design

Utilizing the input stream details, the study designs suitable RPB units employing a heuristics-based approach. The primary design variables of an RPB unit include the inner radius ($r_{inner}$), outer radius ($r_{outer}$), and packing height ($H$). In an RPB, the radii are analogous to the height of a traditional PB due to radial flow, while the height corresponds to the diameter, influencing the unit capacity and pressure drop.

The design process hinges on analyzing flooding conditions like conventional packed beds. Since the swiftest vapor velocity in an RPB occurs near the inner radius, ensuring this area is free from flooding is a critical design consideration. To determine the optimal RPB design parameters, we utilized a systematic RPB design procedure proposed by Agarwal et al. [10]. This process acknowledges the need for a compact inner radius while mitigating the risks of excessive exit vapor velocity, which can lead to flooding and jet formation. The formula for the minimum inner radius, balancing compactness with acceptable exit velocity is given as:

$$r_{inner,min} = \left(\frac{G}{\pi u_{jet}(1 - f_d)}\right)^{1/2} \left(\frac{\rho^V p}{\rho^L}\right)^{1/4} \tag{14}$$

For a specified inner radius, the minimum height of the RPB required to avoid flooding, referred to as the flooding height ($H_{flood}$), can be calculated using the following equation:

$$H_{flood} = \frac{G}{2\pi r_{inner} u^V_{design}} \tag{15}$$

Here, $u^V_{design}$ represents the superficial vapor velocity designated for RPB design, typically a fraction of the flooding vapor velocity, such as 80% used in this study. To determine the flooding vapor velocity, we employ a modified Sherwood dimensionless analysis specifically tailored for RPB units, as developed by Jassim [12]:

$$ln\left[\frac{{u^V_{flood}}^2 a_p}{r\omega^2\varepsilon^3}\left(\frac{\rho^V}{\rho^L}\right)\right] = -3.01 - 1.40\, ln\left(\frac{L}{G}\sqrt{\frac{\rho^V}{\rho^L}}\right) - 0.15\left[ln\left(\frac{L}{G}\sqrt{\frac{\rho^V}{\rho^L}}\right)\right]^2 \tag{16}$$

This approach allows for a more accurate estimation of the critical height to prevent flooding in the RPB, ensuring a reliable and efficient design process.

The internal flooding conditions of RPB units are defined as the ratio of the internal vapor velocity to the flooding vapor velocity as follows:

$$\phi_{flood} = \frac{u^V}{u^V_{flood}} \tag{17}$$

The outer radius of the RPB unit is determined through a differential material balance focused on the primary component, $CO_2$. This process involves numerically integrating from the inner to the radius, ensuring that the degree of separation meets the specified requirements. The calculation is formulated as follows:

$$2\pi\int_{r_{inner}}^{r_{outer}} a(r)\cdot r\cdot dr = \frac{V}{H\cdot K^{overall}_{CO_2}}\int_{y_{in,CO_2}}^{y_{out,CO_2}} \frac{dy}{\left(y_{CO_2} - y^*_{CO_2}\right)} \tag{18}$$

Here, $a(r)$ represents the effective specific surface area, which varies with $r$ due to the acceleration term. For this study, an outer radius calculation assumes a 90% removal target ($y_{out,CO_2} = 0.1 y_{in,CO_2}$).

Once the RPB unit configurations are established, evaluating the pressure drop within units is critical to ensure the design's mechanical viability. The pressure drop across the packing zone can be approximated using an integrated model [10]:

$$\Delta P_{packing} = \frac{f\rho_g}{2d_h}\left(\frac{G}{2\pi H}\right)^2\left(\frac{1}{r_{inner}} - \frac{1}{r_{outer}}\right) + \frac{\rho_g}{2}\left(\frac{G}{2\pi H}\right)^2\left(\frac{1}{r_{inner}^2} - \frac{1}{r_{outer}^2}\right) + \frac{\rho_g}{2}A\omega^2\left(r_{outer}^2 - r_{inner}^2\right) \quad (19)$$

To optimize the efficiency and cost-effectiveness of an amine-based $CO_2$ capture process, it is advisable to keep the pressure drop within a reasonable range, mindful of the costs related to the inlet stream blower or compressor. An excessive pressure drop could require modifications to the RPB system's design to maintain operational viability. This may involve modifying the inner radius, the height of the unit, or operating conditions, particularly the rotation speed. An initial estimate of the appropriate operating conditions can be derived using the operating loading capacity and a straightforward material balance on $CO_2$:

$$F^V = \frac{F^L y_{CO_2}\eta_{CO_2}}{(\alpha_{Rich} - \alpha_{Lean})x_{MEA}} \quad (20)$$

Here, $F^V$ and $F^L$ represent the molar flow rates of the inlet flue gas and lean amine solvent, respectively, and $\eta_{CO_2}$ is the $CO_2$ capture efficiency. The study assumes the use of stainless steel wire mesh for the RPB units with parameters from Cheng et al. [39] and sets the rotation speed at 600 rpm based on the previous 1 TPD scale of the pilot plant investigations [18]. An engineering factor of 1.3 is also assumed for the vapor flow rate. Table 3 outlines the assumed process variables and details the RPB absorber and stripper designs for various MEA concentrations. Notably, while the inner radius remains relatively constant due to its reliance on gas input conditions, the outer radius and height tend to decrease with increasing MEA concentration. This trend is linked to the lower L/G ratio and enhanced absorption rate at higher MEA concentrations, demonstrating the influence of solvent concentration on the RPB design and efficiency.

Table 3 Process variable assumption and RPB design (sequential design)

| Variable type | Variables | 30wt% | 50wt% | 70wt% |
|---|---|---|---|---|
| Process variable assumptions | L/G ratio (kg/kg) | 3.87 | 2.39 | 1.76 |
| | Lean/Rich loading (mol/mol) | | 0.23/0.50 | |
| | Rotation speed (RPM) | | 600 | |
| | Flooding ratio (%) | | 80 | |
| | Packing type (-) | | Stainless steel wire mesh | |
| | Specific surface area ($m^2/m^3$) | | 803 | |
| | Packing void fraction (-) | | 0.96 | |
| RPB design (Absorber) | Inner radius (m) | 0.18 | 0.18 | 0.18 |
| | Outer radius (m) | 1.31 | 1.28 | 1.19 |
| | Height (m) | 0.36 | 0.30 | 0.28 |
| RPB design (Stripper) | Inner radius (m) | 0.10 | 0.10 | 0.09 |
| | Outer radius (m) | 0.35 | 0.27 | 0.23 |
| | Height (m) | 0.43 | 0.36 | 0.33 |

#### 3.2.2. Operating condition optimization

Once the design parameters for the RPB units are established, it is crucial to identify an energy-efficient set of operating conditions for the overall process in the sequential process design approach. The RPB-based $CO_2$ capture process consumes both heat for solvent regeneration and electricity for RPB rotation. To simplify the optimization problem and simultaneously account for these two forms of energy consumption, we employ a conversion factor of 0.4, reflective of a typical thermal power plant's efficiency [42]. This conversion factor aligns with the expected ratio of electricity to steam energy costs, which will be further explored in the Techno-Economic Analysis (TEA) as detailed in Table 4. This approach ensures a consistent and rational basis for evaluating the energy requirements of the RPB-based $CO_2$ capture process. The objective is to minimize the total energy consumption for $CO_2$ capture, formulated as follows:

$$\min_{x} E_{Cap} = E_{SRD} + E_{SRE} \tag{21}$$

where $E_{SRD} = \frac{Q_{Reb}}{\dot{m}_{CO_2,Cap}}$ & $E_{SRE} = \frac{1}{C_{Elec_to_heat}} \frac{E_{Abs}^{rotation} + E_{Str}^{rotation}}{\dot{m}_{CO_2,Cap}}$

$$x \in [F_{Solv}, T_{Reb}, P_{Str}, \omega_{Abs}, \omega_{Str}] \quad (22)$$

Here, $E_{Cap}$, $E_{SRD}$ and $E_{SRE}$ represent the total specific $CO_2$ capture energy, specific reboiler duty (SRD) and specific rotation energy (SRE), respectively. The decision variables in this optimization problem include the recirculated solvent flow rate ($F_{Solv}$), the reboiler temperature ($T_{Reb}$), the stripper pressure ($P_{Str}$), and the rotation speeds for both RPB units ($\omega_{Abs}$, $\omega_{Str}$). These variables were chosen for their significant impact on the process's efficiency, impacting the carbon capture rate and energy requirements [18]. The energy required to rotate the RPB units, denoted as $E_{RPB}^{rotation}$, can be calculated using the equation from Singh et al. [43]:

$$E_{RPB}^{rotation} = 1.2 + 1.1 \times 10^{-3} \rho^L L r_{outer}^2 \omega^2 \quad (23)$$

This equation accounts for the liquid density ($\rho^L$), the volumetric liquid flow rate ($L$), the outer radius of the RPB unit ($r_{outer}$), and the rotation speed ($\omega$). The optimization process also incorporates several critical constraints to ensure the system operates within practical and safe parameters:

$$f_{process}(x, para) = 0 \quad (24)$$

$$\eta_{CO_2} \geq 90\% \quad (25)$$

$$T_{Reb} \leq 120°C \quad (26)$$

$$0\% \leq \phi_{flood,RPB} \leq 80\% \quad (27)$$

The first constraint states the steady-state version of the developed process model ($f_{process}$) needs to be satisfied, while the second constraint ensures a minimum 90% $CO_2$ capture rate is achieved. To prevent the thermal degradation of the MEA solvent, the reboiler temperature is capped at 120°C. Finally, the flooding ratio for each RPB unit must stay within these bounds to ensure operation without flooding. To tackle this complex optimization problem, the study employs the NLPMSO (nonlinear programming with multi-starting points) global optimization solver in gPROMS. This solver is adept at finding the optimal set of operating conditions that meet the energy efficiency targets and adhere to the specified constraints, ensuring the RPB-based $CO_2$ capture process is both effective and practical.

### 3.3. Simultaneous design approach

#### 3.3.1. Limitations of the sequential design approach

In the sequential design approach for RPB units, accurately assuming operating conditions is crucial. While liquid and vapor flow rates are key parameters for traditional PBs, the RPB design requires an additional critical parameter, the packing rotation speed. The rotation speed significantly influences the RPB's hydrodynamics, including liquid flow patterns, holdup, surface area, and particularly the flooding point [44]. As such, it profoundly affects both process performance and the physical design of the RPB, including its susceptibility to flooding. However, due to the interconnected nature of these variables in RPB design, the sequential design approach is inherently limited to identifying local optima rather than the absolute best solution for RPB-based processes. Here, we advocate the simultaneous design approach to address this limitation.

#### 3.3.2. Problem formulation

In the simultaneous design approach, which concurrently determines the RPB design and operating conditions, an optimization problem is constructed to minimize the total annual cost (TAC) per captured $CO_2$ ($\dot{m}_{CO_2,Cap}$). In contrast to the earlier sequential methodology, which first establishes RPB design based on predefined operational conditions and subsequently refines these conditions, our approach simultaneously considers both RPB design parameters ($d$) and and operational variables ($x$) as decision variables:

$$\min_{d,x} C_{Cap} = \frac{TAC}{\dot{m}_{CO_2,Cap}} = \frac{ACC + AOC}{\dot{m}_{CO_2,Cap}} \tag{28}$$

The decisions variables ($d$) for the RPB design include:

$$d \in \left[r_{inner,Abs}, r_{outer,Abs}, H_{Abs}, r_{inner,Str}, r_{outer,Str}, H_{Str}\right] \tag{29}$$

And the same set of operating variables ($x$) are considered as:

$$x \in [F_{Solv}, T_{Reb}, P_{Str}, \omega_{Abs}, \omega_{Str}] \tag{30}$$

Here, $ACC$ represents the annualized capital cost, and $AOC$ the operating cost. The capital cost is estimated using the Lang factor method, which applies a multiplier to the free-on-board (FOB) purchase cost of the equipment. Meanwhile, the operating cost is calculated based on the energy consumption determined from the process simulation. The objective is to find the combination of RPB design and operating conditions that results in the lowest possible $CO_2$ capture cost ($C_{Cap}$), balancing upfront investment with ongoing expenses. By integrating the determination of RPB design and operational conditions, this simultaneous approach anticipates achieving a global optimum in costs within the established constraints, moving beyond the limitations of the sequential method.

The annualized capital cost is calculated using the following formula:

$$ACC = \left(\frac{i(i+1)^n}{(i+1)^n - 1}\right) \times CAPEX \tag{31}$$

Here, CAPEX is the capital expenditure determined by:

$$CAPEX = 5.93 \times 1.05 \times \sum_{i \in \text{devices}} \left(\frac{816}{567}\right) C_{FOB,i} \tag{32}$$

This calculation employs the Lang factor of 5.93 for fluid processing processes [45], and adjusts for inflation using the 2022 Chemical Engineering Plant Cost Index (CEPCI) of 816 (with a base of 567 in 2013). A 20-year plant lifetime and a 10% interest rate are assumed. Given the absence of a specific capital cost model for commercial-scale RPB units, we adopt a model designed for centrifuges, a common practice for estimating the costs of rotating devices [46, 47]. The FOB purchase cost for the RPB unit includes costs for rotation components and packing bed, modeled after a large-scale vertical auto-batch centrifuge [48]:

$$C_{RPB}^{FOB} = C_{Centrifuge}^{FOB} + V_{RPB}^{Packing} C_{RPB}^{Packing} \tag{33}$$

$$C_{Centrifuge}^{FOB} = \$6180 \cdot (D_{outer})^{0.94} \tag{34}$$

Here, $V_{RPB}^{Packing}$ and $C_{RPB}^{Packing}$ represent the volume and cost of packing, respectively, with a packing price of \$285/ft$^3$ and $D_{outer}$ is the centrifuge's diameter in inches. The detailed assumptions and FOB purchase cost models for other process units can be found in the Supplement materials.

The annualized operating cost *(AOC)* is formulated as:

$$AOC = 1.1 \times \left(C_{Utility} + C_{M\&O} + C_{Overhead} + C_{Material}\right) \tag{35}$$

$$C_{Utility} = \frac{330}{365}\left(C_{Steam}Q_{Reb} + C_{Elec}\left(E_{RPB}^{rotation} + E_{Comp} + E_{Pump}\right) + C_{CW}Q_{Cool}\right) \tag{36}$$

Utility costs are sourced from Chung and Lee [49], with 10% general expenses and 330 operation days considered, and summarized in Table 4. Here, we refer to the summation of M&O and overhead costs as fixed cost. The details for other cost terms can be found in the Supplement materials.

Table 4 Cost for various utility types

| Utility type | Variables | Price | Unit |
|---|---|---|---|
| Steam | $C_{Steam}$ | 8.0 | \$/GJ |
| Electricity | $C_{Elec}$ | 19.2 | \$/GJ |
| Cooling water | $C_{Cool}$ | 0.015 | \$/GJ |

Additionally, the optimization includes constraints on the process and RPB design:

$$f_{process}(x, para) = 0 \tag{37}$$

$$\eta_{CO_2} \geq 90\% \tag{38}$$

$$T_{Reb} \leq 120°C \tag{39}$$

$$0\% \leq \phi_{flood,RPB} \leq 80\% \tag{40}$$

$$r_{inner,min,j} \leq r_{inner,j} \leq r_{outer,j} \tag{41}$$

$$\frac{H_j}{2r_{outer,j}} \leq 0.5 \tag{42}$$

The index $j$ represents the RPB units, encompassing absorber ($Abs$) and stripper ($Str$) columns. To integrate existing heuristic knowledge, mechanical constraints have been introduced as bounds for RPB inner and outer radii, with the minimum inner radius ($r_{inner,min}$) calculated based on Eq. (14). The last constraint ensures adhering to design recommendations for large-scale RPB units, as proposed by Trent [50]. The optimization formulation allows for the incorporation of other design criteria or requirements as needed. The optimization aims to achieve optimal design and operating conditions that minimize the total $CO_2$ capture cost while meeting all process requirements and mechanical constraints. The NLPMSO global optimization solver is utilized to navigate this complex landscape and avoid local optima.

## 4. Results and Discussion

### 4.1. Techno-economic analysis

A plant-wide model for $CO_2$ capture process using standard fixed PBs in both absorber and stripper columns has also been developed for comparison. This benchmark process uses a 30wt% MEA solvent, and its column designs are based on the Generalized Pressure Drop Correlation (GPDC) method [51], assuming 80% flooding conditions. The operating conditions are determined through an optimization problem focused solely on minimizing the SRD, like Eq. (21).

#### 4.1.1. Cost evaluation with sequential design approach

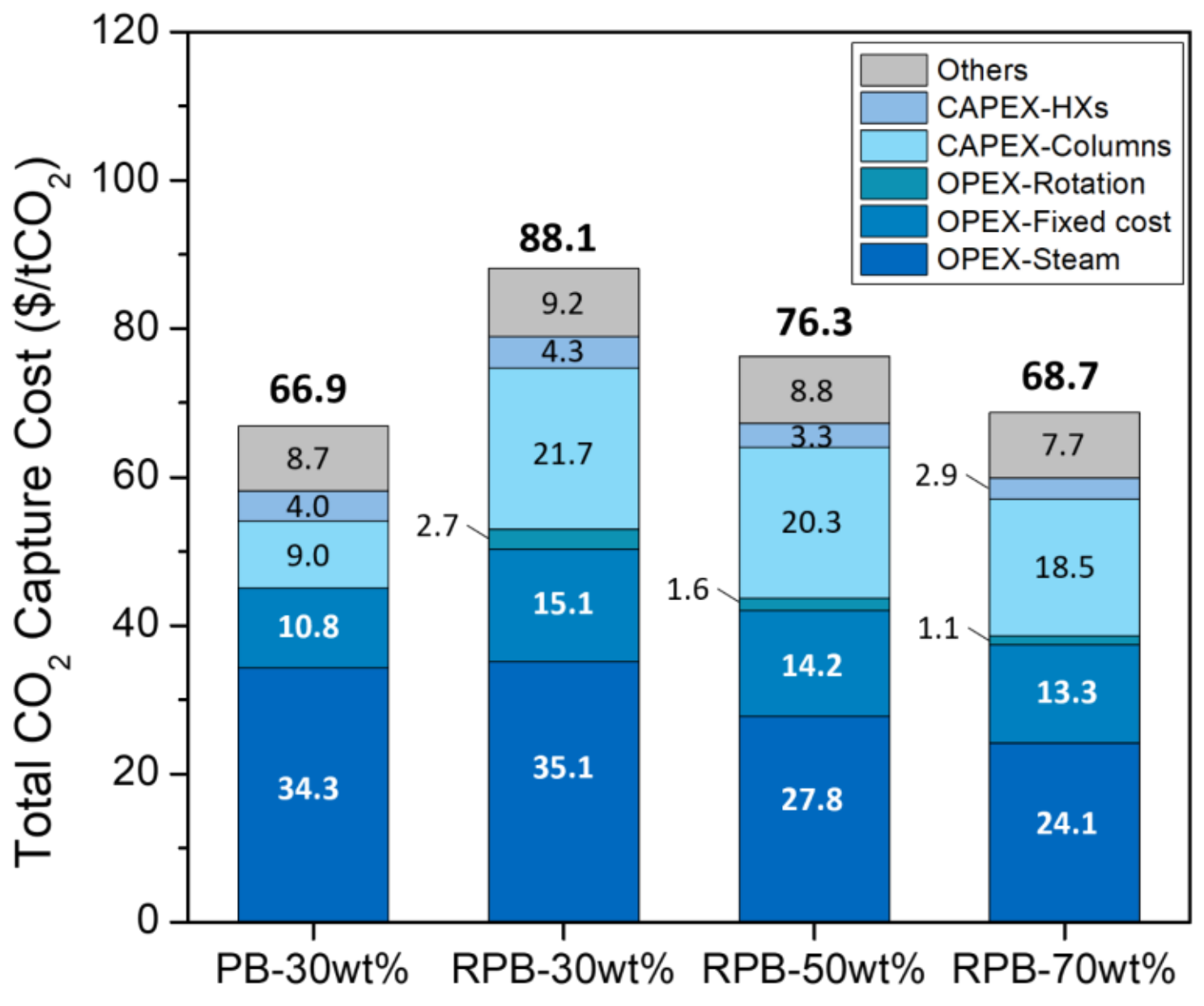

Figure 7 Evaluation of costs on RPB-based $CO_2$ capture processes using the sequential design approach

Figure 7 presents the total $CO_2$ capture cost and its detailed components based on MEA concentration, utilizing the sequential design approach. It is evident that capture costs using RPB columns exceed those with conventional PBs. The primary factor for this increased expense is the capital cost of RPB units, constituting approximately 25% of the total capture cost (or roughly 65% of the capital cost). Conversely, the costs for columns with fixed PBs represent only 14% of the total (or 46% of the capital cost).

Section 4.2.1 will further discuss how RPB columns significantly decrease the packing volume. Yet, the capital costs for RPB units remain considerably higher than those for standard PBs. This trend has been observed in previous RPB applications in the aroma absorption process [47], although some studies have reported potential capital cost reductions with RPB units [19, 46]. The wide variation in these findings might stem from the lack of a uniform commercial-scale RPB capital cost model and inherent high uncertainty in cost estimations. For instance, using Otitoju's cost model [19] can result in a 63% to 80% decrease in RPB capital costs and a total capture cost of $69.6-48.4 GJ/tCO_2$, which is 72% to 104% of the total cost from the reference PB system (Table 5). However, this study aims to identify general trends rather than specific numbers, considering the lack of practical experience with large-scale RPB implementations. Therefore, a conservative and high-cost model, based on similar commercial-scale process units, is employed for cautious cost estimations.

Table 5 Techno-economic analysis results with different RPB capital cost models

| | PB | RPB cost model (this study) | | | RPB cost model from [19] | | |
|---|---|---|---|---|---|---|---|
| | 30wt% | 30wt% | 50wt% | 70wt% | 30wt% | 50wt% | 70wt% |
| Capital cost (k$/yr) | 606 | 1,023 | 945 | 869 | 603 | 467 | 407 |
| -RPB capital cost (k$/yr) | 281 | 676 | 635 | 578 | 256 | 157 | 115 |
| Fixed cost (k$/yr) | 337 | 472 | 443 | 415 | 316 | 265 | 243 |
| Utility cost (k$/yr) | 1,141 | 1,253 | 990 | 858 | 1,253 | 990 | 858 |
| Total capture cost (k$/yr) | 2,085 | 2,748 | 2,378 | 2,142 | 2,171 | 1,722 | 1,508 |
| Total capture cost ($/tCO_2$) | 66.9 | 88.1 | 76.3 | 68.7 | 69.6 | 55.2 | 48.4 |

The cost increase due to the rotation of the packing seems marginal, with the associated electricity costs contributing a minor portion, approximately 2-3% of the total cost. It is significant to note the decrease in total cost as MEA concentrations rise. Concentrated MEA solvent enables the attainment of desired separation efficiency with a lower solvent flow rate and a smaller column packing size, thanks to its enhanced absorption capacity and reaction rate. Moreover, the higher MEA concentration is beneficial for SRD as it minimizes energy loss related to the evaporation of water in the stripping sections. While acknowledging the constraints of thermodynamic limits, the primary benefits of RPB units are their compact packing size and a wide range of solvent options. In particular, the extensive solvent selection available with RPB has the potential to lower both energy consumption and overall cost.

### 4.1.2. Cost evaluation with simultaneous design approach

Figure 8 illustrates the cost savings achievable in RPB-based $CO_2$ capture processes when employing the simultaneous design approach. This method significantly reduces the total capture costs by 9.4% to 12.7% compared to the sequential approach for various MEA concentrations. The majority of these savings come from lower capital costs of the RPB columns and reduced fixed maintenance and operating (M&O) or overhead expenses, which are derived from the CAPEX. Additionally, the simultaneous design method leads to a decrease in the rotation energy costs, although there is a minor increase in the steam energy costs. Notably, even with a conservative and higher-cost model for RPB capital costs, employing MEA concentrations above 50wt% results in capture costs that are competitive with those of a conventional packed bed using 30wt% MEA.

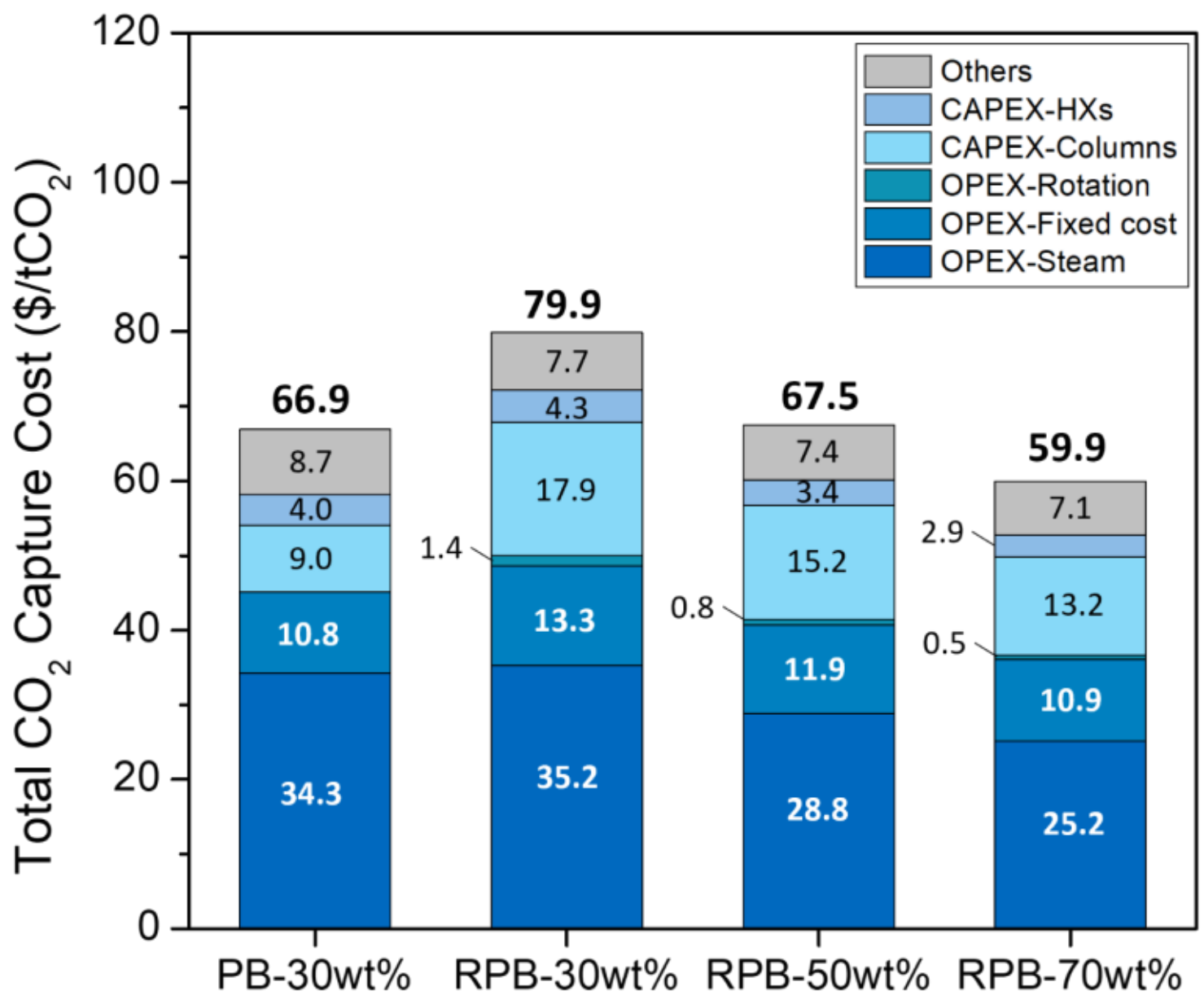

Figure 8 Evaluation of costs on RPB-based $CO_2$ capture processes using the simultaneous design approach

### 4.1.3. Sensitivity analysis

A local sensitivity analysis was conducted to assess how the economic parameters affect the total $CO_2$ capture costs. Figure 9 shows the influence of varying each economic parameter within a ±20% range on the total $CO_2$ capture costs for both RPB and PB processes, using the process design and operating conditions determined from the simultaneous design approach for the RPB-based process.

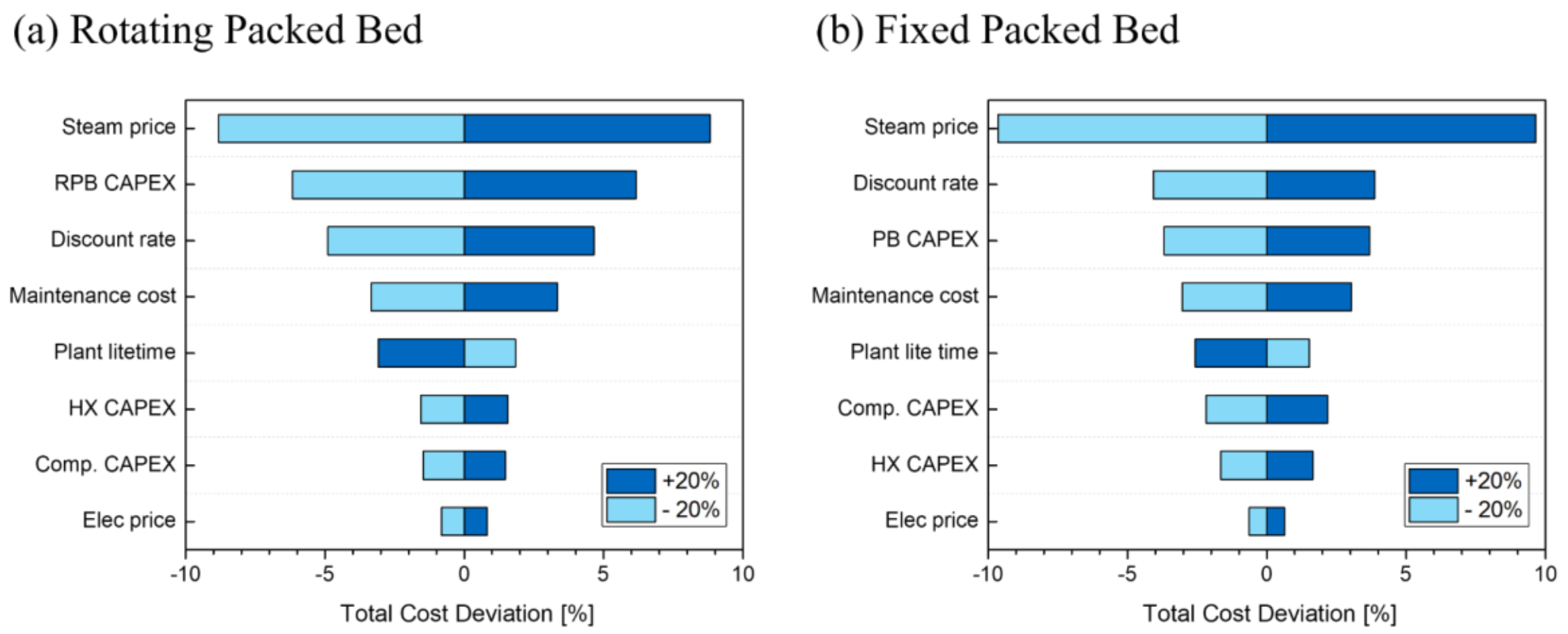

Figure 9 Sensitivity on economic parameters (30wt% MEA)
(a) Rotating Packed Bed (left); (b) Fixed Packed Bed (right)

In both scenarios, the analysis reveals that steam price has the most significant impact. The steam energy cost consistently accounts for the largest portion of the total costs (as seen in Figure 8), making it highly sensitive to changes in steam prices. Following steam price, the capital cost of RPB and PB, or discount rate, are the next most influential factors on total capture costs. It is important to note that despite the packing column's significant impact on overall costs, there is currently no accurate estimation model of RPB capital cost. The high uncertainty in cost estimation can affect the total cost significantly. Other economic parameters showing considerable sensitivity are the maintenance cost and plant lifetime, in that order.

### 4.2. RPB design and operating condition

While cost evaluation offers valuable insights into the economic viability of RPB-based $CO_2$ capture processes, conducting an in-depth technical evaluation is equally crucial to understand the origins of cost savings and gather technical insights for scale-up. Hence, examining the optimal decision variables obtained from the optimization results is key to identifying the most efficient strategies for large-scale RPB-based $CO_2$ capture processes. Table 6 shows the optimized design and/or operating conditions derived from both sequential and simultaneous approaches to RPB design and process operation decisions.

Table 6 Process design parameters and operating conditions according to MEA concentration and design approaches (*: constraint active)

| Category | Variables | 30wt% MEA | | | 50wt% MEA | | 70wt% MEA | |
|---|---|---|---|---|---|---|---|---|
| | | Packed bed | Sequen-tial | Simul-taneous | Sequen-tial | Simul-taneous | Sequen-tial | Simul-taneous |
| Column design | ABS inner radius (m) | - | 0.182 | 0.182* | 0.182 | 0.182* | 0.181 | 0.181* |
| | ABS outer radius (m) | 0.95 | 1.308 | 1.032* | 1.277 | 0.909* | 1.190 | 0.796* |
| | ABS height (m) | 10 | 0.361 | 1.032* | 0.304 | 0.909* | 0.276 | 0.796* |
| | STR inner radius (m) | - | 0.100 | 0.100* | 0.098 | 0.098* | 0.096 | 0.094* |
| | STR outer radius (m) | 0.55 | 0.351 | 0.275* | 0.286 | 0.209* | 0.231 | 0.179* |
| | STR height (m) | 5 | 0.425 | 0.275* | 0.357 | 0.209* | 0.331 | 0.179* |
| Operating condition | Solvent flowrate (kg/s) | 19.5 | 21.2 | 21.7 | 12.5 | 13.8 | 9.1 | 10.0 |
| | Reboiler temperature (°C) | 120* | 120* | 120* | 120* | 120* | 120* | 120* |
| | Stripper pressure (kPa) | 1.86 | 1.86 | 1.88 | 1.58 | 1.62 | 1.08 | 1.13 |
| | ABS rotation speed (rpm) | - | 490 | 267 | 498 | 279 | 502 | 292 |
| | STR rotation speed (rpm) | - | 503 | 786 | 415 | 756 | 425 | 888 |
| Process variables | $CO_2$ capture rate (%) | 90* | 90* | 90* | 90* | 90* | 90* | 90* |
| | L/G ratio (kg/kg) | 3.34 | 3.63 | 3.71 | 2.15 | 2.36 | 1.56 | 1.71 |
| | Lean loading (mol/mol) | 0.23 | 0.23 | 0.24 | 0.23 | 0.24 | 0.22 | 0.23 |
| | Rich loading (mol/mol) | 0.50 | 0.48 | 0.49 | 0.49 | 0.48 | 0.48 | 0.47 |
| | ABS pressure drop (kPa) | 2.0 | 2.9 | 0.5 | 3.1 | 0.4 | 3.0 | 0.4 |
| | ABS flooding (%) | 53 | 80* | 52 | 80* | 49 | 80* | 49 |
| | Steam energy (kW) | 4,008 | 4,367 | 4,377 | 3,454 | 3,576 | 3,000 | 3,123 |
| | Rotation energy (kWe) | - | 141.4 | 73.6 | 85.1 | 42.4 | 57.3 | 27.9 |
| Performance indices | SRD (GJ/t$CO_2$) | 3.67 | 3.99 | 4.00 | 3.15 | 3.27 | 2.74 | 2.86 |
| | SRE (GJe/t$CO_2$) | - | 0.129 | 0.067 | 0.078 | 0.039 | 0.052 | 0.026 |
| | Capture energy (GJ/t$CO_2$) | 3.67 | 4.32 | 4.17 | 3.35 | 3.37 | 2.87 | 2.92 |
| | Capture cost ($/t$CO_2$) | 66.9 | 88.1 | 79.9 | 76.3 | 67.5 | 68.7 | 59.9 |

#### 4.2.1. Packing design and volume

The optimization results highlight the well-documented advantage of reduced packing volume in RPB, showing an 8.5 to 14.9 times decrease compared to PB columns with a 30wt% MEA solvent baseline. This significant disparity in packing volumes is illustrated in Figure 10, and becomes even more marked with higher MEA concentrations. Using 70wt% MEA, the packing volume can be reduced by as much as 23.6 times, owing to the synergetic effects of the concentrated solvent's enhanced absorption capacity and the RPB column's design.

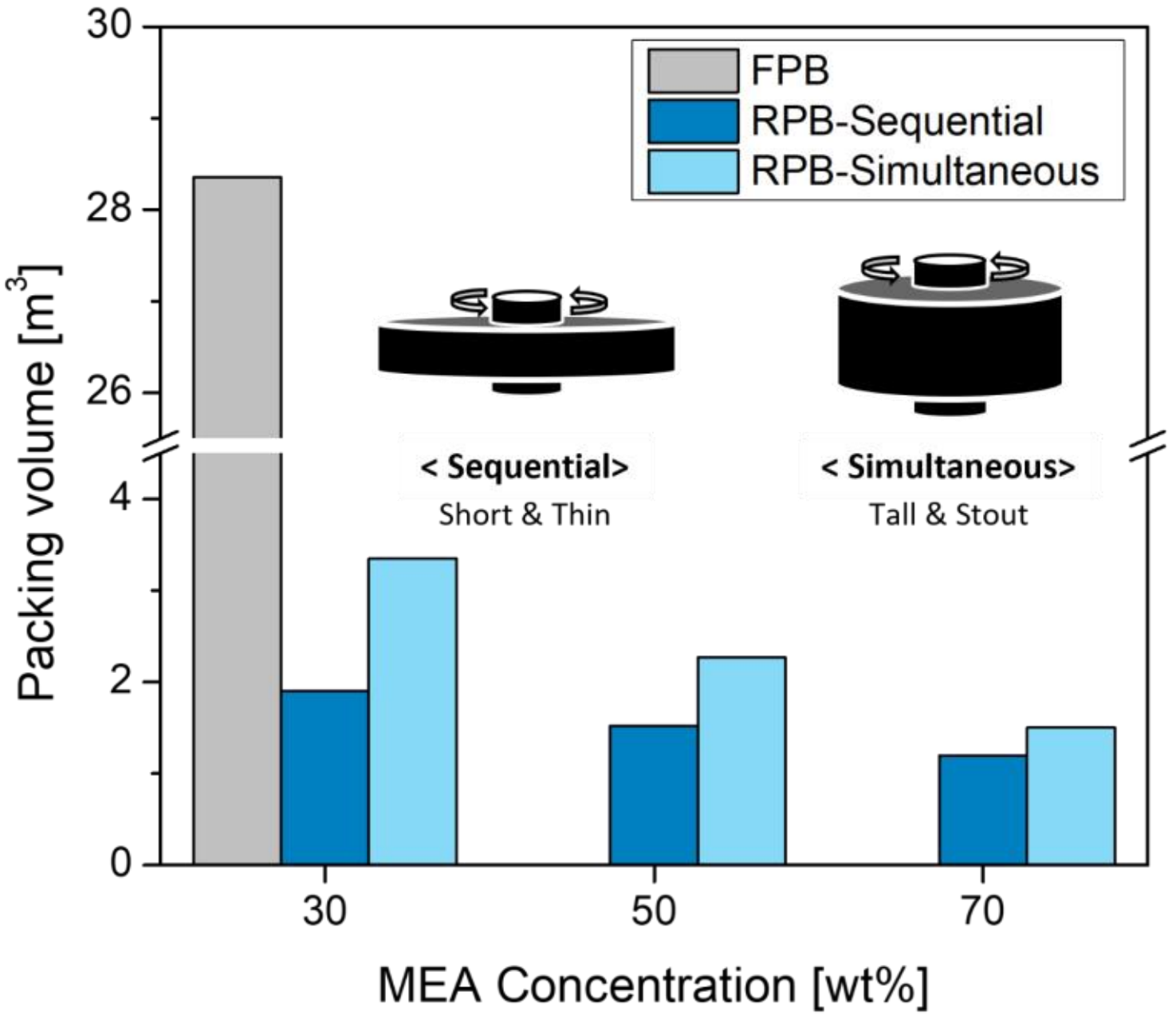

Figure 10 Required packing volume for the absorber column with PB and RPB

Notably, the optimal RPB design varies based on the design approach, as detailed in Table 6. Sequential and heuristic designs tend to favor a shorter, thinner, disk-shaped RPB, while our analysis suggests that a taller, thicker configuration is more cost-effective. Although smaller RPB heights might seem beneficial for reducing capital costs in small-scale processes, as empirically recommended [10], a larger height and reduced radius can decrease both capital and operational costs at larger scales, maintaining equivalent process throughputs. Although the required packing volume for the simultaneous approach appears to be larger than the sequential approach, the resulting capital costs can be lower since the capital costs for RPB units typically correlate with the power consumption or diameter of the RPB. Moreover, as the RPB radius increases, both rotation energy and pressure drop rise quadratically, making a large radius less desirable for large-scale RPBs in terms of operating costs.

It is noteworthy that the design recommendation constraint dictates the outer radius and height share identical values. While a thin disk-shape RPB is commonly favored for lab-scale settings, considerations shift on a larger scale where the influence of pressure drop becomes more pronounced. In such cases, it might be economically advantageous to minimize pressure drop by augmenting the height. This trend appears in the direction of increasing the RPB height as much as possible and requires activating the RPB design recommendation constraint. For practical use, the validation of this stout RPB design, ensuring that the hydraulic condition remains above the loading point is crucial for adequate packing wetting. It is worth noting that these constraints can be readily adjusted to align with practical requirements.

#### 4.2.2. Optimal operating conditions

In amine-based $CO_2$ capture processes, including those with large recycle streams, the relationship between stripping temperature and the equilibrium partial pressure of $CO_2$ suggests that higher stripping temperature can enhance stripping efficiency. However, excessively high temperatures can cause phase changes not just in $CO_2$ but also in $H_2O$. Therefore, the ideal operating conditions aim to find a balance in stripping temperature and pressure that promotes $CO_2$ phase change while minimizing $H_2O$ evaporation. Due to the need to prevent solvent degradation, the optimal reboiler temperature is often set at its maximum allowable level. Consequently, this leads to the activation of the constraint on reboiler temperature, maintaining at 120°C in all scenarios (refer to Table 6). Additionally, it is observed that optimal stripping pressure decreases as MEA concentration increases, which correlated with a reduction in $H_2O$'s partial pressure [18]. Figure 11 illustrates how optimal stripping pressure and L/G ratio vary with different MEA concentrations. As the contribution of $H_2O$ vapor pressure rapidly diminishes with a decreasing portion of $H_2O$, the optimal stripping pressure tends to decrease as MEA concentration increases. Using concentrated MEA solvents allows for achieving the desired capture rate with a lower solvent flow rate due to their absorption capacity, which in turn reduces the L/G ratio. This reduction can also lead to decreased capital costs of other process units.

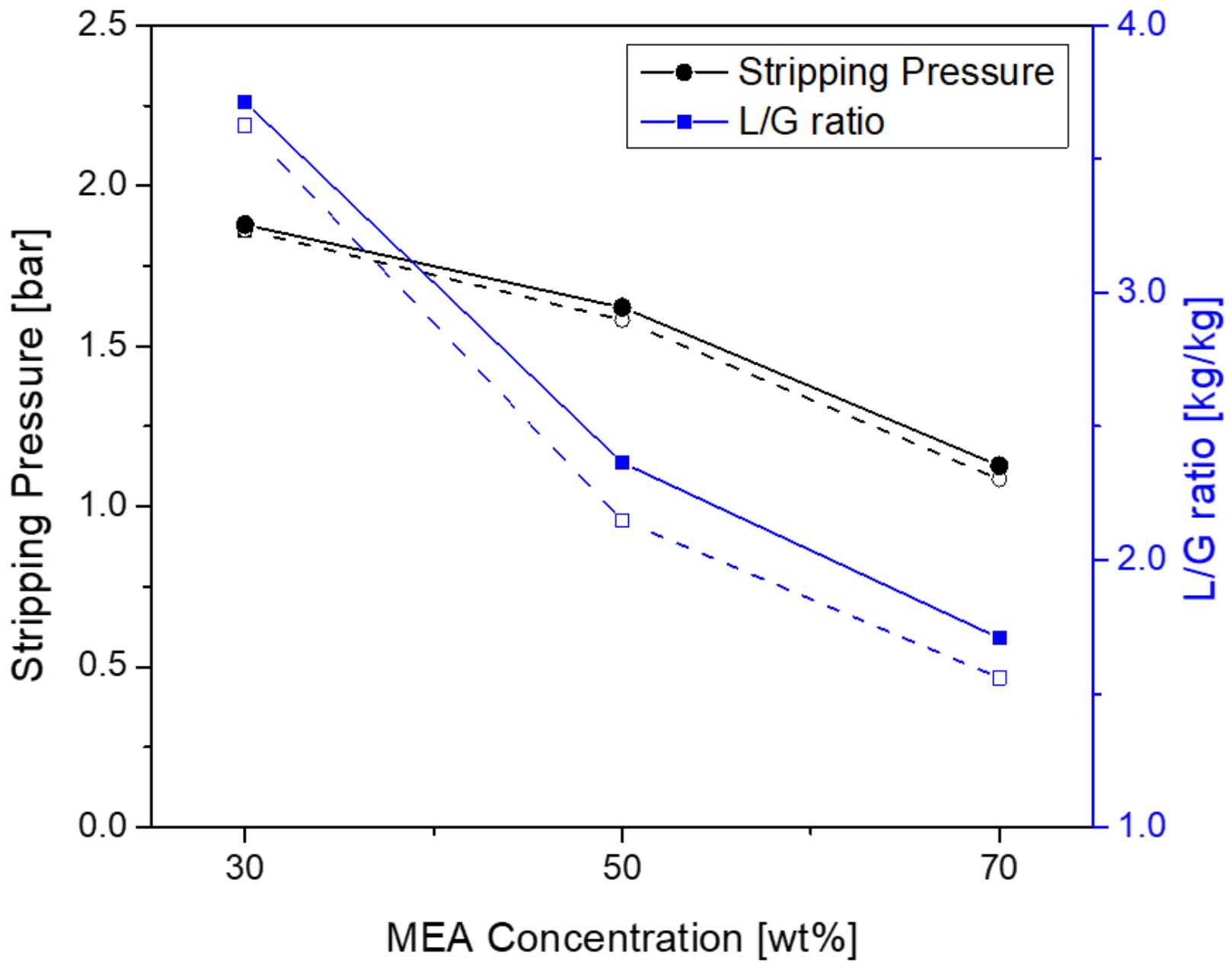

Figure 11 Optimal stripping pressure and L/G ratio with varying MEA concentration (solid line: simultaneous approach; dash line: sequential approach)

Regarding optimal rotation speeds, they show little variation with changes in solvent concentration, but differ significantly between design approaches. For the absorber column, the sequential design suggests a rotation speed of approximately 500 rpm, whereas the simultaneous design recommends 250 to 300 rpm. A critical point to note is the flooding ratio within the absorber columns; the sequential strategy often reaches an upper limit of 80% indicating that the RPB design and hydraulic conditions are somewhat fixed to the assumed 600 rpm during the RPB design stage (as mentioned Table 3). This highlights the interdependency between RPB design and operating conditions, emphasizing that understanding this relationship is key to developing an integrated and cost-effective RPB-based $CO_2$ capture process.

### 4.3. Discussion and perspective

The RPB is a forward-thinking process unit designed to enhance mass transfer rates significantly. Its primary advantage lies in the substantial reduction of packing volume. While the RPB doesn't inherently offer the reduced energy due to the additional electrical energy required for packing rotation, it opens opportunities for using a broader range of solvents, some of which were previously unsuitable for conventional PBs, potentially leading to overall energy savings in $CO_2$ capture processes.

A thorough understanding of RPB units' strengths and weaknesses allows for identifying appropriate applications for this technology. RPB's efficient space utilization greatly expands its industrial applicability and the feasibility of $CO_2$ capture technologies in space-constrained environments. For instance, it makes $CO_2$ capture more viable in areas with spatial limitations and in scenarios previously considered impractical, like onboard $CO_2$ capture and storage (OCCS). The compact nature of RPB also suggests benefits in operational and control flexibility, offering a stark contrast to the slow dynamics of traditional amine-based $CO_2$ capture processes [52, 53].

A pivotal consideration in deploying RPB-based $CO_2$ capture technology is identifying the optimal balance between practicality and economic viability. As we scale up, the costs associated with higher rotational speeds and increased pressure drops may incrementally elevate operational expenses, diverging from traditional methods that predominantly gain from economies of scale. Additionally, the enlargement of process units introduces heightened safety concerns due to the inherent rotation involved, presenting practical constraints [54]. This dynamic suggests the existence of an optimal scale for RPB-based $CO_2$ capture processes, wherein the advantages of economies of scale effectively counterbalance the supplementary costs and practical hurdles linked to the mechanical rotation of the packing. By modularizing the RPB-based $CO_2$ capture process around this ideal scale, we can potentially enhance its economic appeal. Such a strategy permits a phased expansion, facilitating a more strategic approach to optimizing the economic efficiency of $CO_2$ capture operations.

However, analyzing the commercial-scale RPB-based process is challenging due to the lack of precise and consistent capital cost models for RPB units. As the capital cost of the RPB column is a significant contributor to the total expense, developing accurate cost models is crucial for informed decision-making and the advancement of this technology. Further research and practical implementation of large-scale RPB-based $CO_2$ capture processes are necessary to refine cost estimations and understand the full potential and limitations of RPB technology.

## 5. Conclusion

This comprehensive study examines the RPB-based $CO_2$ capture process with MEA solvents through detailed process modeling, validation, optimization of design and operating conditions, and economic analysis. The implemented process model, integrating eNRTL and enhancement factor models, demonstrated high prediction accuracy, with a MARE of 11.4% for the $CO_2$ capture rate and 9.9% for energy consumption, benchmarked against pilot plant data.

Our research evaluated the performance and cost of RPB with 30, 50, and 70wt% MEA solvents for a practical-scale application targeting 100 TPD $CO_2$ capture. Addressing the lack of established design procedures for commercial-scale RPB, we proposed a method optimizing both design and operating conditions to minimize total $CO_2$ capture costs. Our findings show that the optimization-led simultaneous design method can significantly reduce costs by 9.4% to 12.7%, resulting in a total cost of \$59.9/t$CO_2$ to \$79.9/t$CO_2$. Notably, the simultaneous design approach indicates that bulkier (i.e., taller and smaller in radius) RPB units might be more suitable for commercial-scale processes due to factors like increased pressure drop, rotation energy, and capital costs. Our simultaneous process design method enables the integration of cost-effective design while meeting heuristic requirements, a crucial aspect for exploring expanded plant scales and operating conditions. This comprehensive approach provides a more nuanced understanding of RPB unit design for commercial-scale $CO_2$ capture, offering valuable insights for future implementations.

Comparing RPB with a 30wt% MEA and fixed PB process highlights RPB's advantages and supports its commercial-scale implementation. Key benefits include a dramatic reduction in packing volume (8.5 to 23.6 times) and the ability to utilize a broader range of solvents. Our study illustrates that significant energy consumption reductions are primarily due to the expanded solvent selection. This implies that developing solvents tailored for RPB could further enhance the efficiency of RPB-based $CO_2$ capture processes. Additionally, we observed that the increased costs due to packing rotation are relatively minor.

A significant challenge in evaluating commercial-scale RPB processes lies in the difficulty of accurately estimating RPB capital costs. These costs are one of major contributor to total expenses, just after steam costs, and are highly sensitive to overall costs. The lack of large-scale implementation experience complicates accurate capital cost predictions. Nonetheless, continued interest and the adoption of large-scale processes are anticipated to overcome these barriers, advancing the understanding and efficiency of RPB technology.

## Acknowledgment

This work was supported by the Korea Institute of Energy Technology Evaluation and Planning (KETEP) funded by the Ministry of Trade, Industry and Energy (MOTIE) (No. 20224C10300050)

## Nomenclature

Variables

| | |
|---|---|
| $a$ | Effective surface area ($m^2/m^3$) |
| $a_p$ | Specific surface area ($m^2/m^3$) |
| $C_{cavity}$ | Correction factor (-) |
| $C_p$ | Molar heat capacity (kJ/mol/K) |
| $C_i$ | Concentration of component i ($mol/m^3$) |
| $C_{tot}$ | Total concentration ($mol/m^3$) |
| $D_i$ | Diffusivity of component i in mixture ($m^2/s$) |
| $d_p$ | Hydraulic diameter (m, $d_p = 4\varepsilon/a_p$) |
| $E$ | Specific energy consumption ($GJ/tCO_2$) |
| $E_{CO_2}$ | Enhancement factor of $CO_2$ (-) |
| $F$ | Molar flow rate (mol/s) |
| $G$ | Volumetric gas flowrate ($m^3/s$) |
| $H$ | RPB height (m) |
| $\Delta H_{H_2O}^{vap}$ | Heat of evaporation of water (kJ/mol) |
| $\Delta H_{CO_2}^{abs}$ | Heat of absorption of $CO_2$ (kJ/mol) |
| $h^I$ | Overall heat transfer coefficient ($kW/m^2/K$) |
| $Ha$ | Hatta number |
| $He_i$ | Henry constant of component i ($Pa{\cdot}m^3/mol$) |
| $K_i^{overall}$ | Overall mass transfer coefficient of component i ($mol/m^2/s/Pa$) |
| $k_{app}$ | Apparent reaction kinetics (1/s) |
| $k_i$ | Mass transfer coefficient of component i (m/s) |
| $L$ | Volumetric liquid flowrate ($m^3/s$) |
| $\dot{m}_{CO_2,Cap}$ | Mass flow rate of capture $CO_2$ (kg/s) |
| $N_i$ | Molar flux of component i ($mol/m^2/s$) |
| $P_i$ | Partial pressure of component i (Pa) |
| $P_i^*$ | Equilibrium partial pressure of component i (Pa) |
| $Q_{Reb}$ | Reboiler duty (kJ/s) |
| $r$ | Radial domain (m) |
| $r_{inner}$ | Inner radius (m) |
| $r_{outer}$ | Outer radius (m) |
| $T$ | Temperature (K) |
| $u$ | Superficial velocity (m/s) |
| $x_i$ | Liquid molar fraction of component i (-) |
| $y_i$ | Vapor molar fraction of component i (-) |

Greek

| | |
|---|---|
| $\alpha$ | $CO_2$ loading (mol $CO_2$/mol Amine) |
| $\rho$ | Mass density (kg/m$^3$) |
| $\gamma$ | Contact angle (deg) |
| $\phi_{flood}$ | Flooding ratio (%) |
| $\eta_{CO_2}$ | $CO_2$ capture rate (%) |
| $\mu$ | Dynamic viscosity (Pa·s) |
| $\varepsilon$ | Fraction of void or holdup (-) |
| $\omega$ | Angular velocity (rad/s) |
| $v$ | Kinematic viscosity (m$^2$/s) |

Abbreviations

| | |
|---|---|
| $Abs$ | Absorber |
| $Cap$ | Capture |
| $Comp$ | Compressor |
| $CW$ | Cooling water |
| $Elec$ | Electricity |
| $FOB$ | Free on board |
| $Reb$ | Reboiler |
| $SRD$ | Specific reboiler duty |
| $SRE$ | Specific rotation energy |
| $Solv$ | Solvent |
| $Str$ | Stripper |